\documentclass[twocolumn,pra,aps]{revtex4}

\usepackage{mathptmx}
\usepackage{subfigure}
\usepackage{psfrag,graphicx}
\usepackage{dcolumn}
\usepackage{amsmath,amssymb}
\usepackage{bm}
\usepackage{color}
\usepackage{latexsym}
\usepackage{epstopdf}
\usepackage{color}
\usepackage[english]{babel}
\usepackage{latexsym}
\usepackage{psfrag,graphicx}
\usepackage{subfigure}
\usepackage{amsmath}
\usepackage{amssymb}
\usepackage{amsfonts}
\usepackage{bm}
\usepackage{natbib}
\usepackage{epstopdf}
\usepackage{appendix}

\definecolor{mygrey}{gray}{0.35}
\definecolor{myblue}{rgb}{0.2,0.2,0.8}
\definecolor{myzard}{cmyk}{0,0,0.05,0}
\definecolor{mywhite}{rgb}{1,1,1}
\definecolor{mywhite}{rgb}{1,1,1}
\definecolor{myred}{rgb}{1,0.,0.3}

\usepackage[colorlinks=true,citecolor=myblue,linkcolor=myred]{hyperref}

\def\ba{\begin{align}}
\def\enda{\end{align}}
\def\bi{\begin{itemize}}
\def\ei{\end{itemize}}

\def\be{\begin{equation}}
\def\ee{\end{equation}}
\def\bea{\begin{eqnarray}}
\def\eea{\end{eqnarray}}
\def\bse{\begin{subequations}}
\def\ese{\end{subequations}}

\begin{document}
\title{Enhanced Two-Parameter Phase-Space-Displacement Estimation Close to Dissipative Phase Transition}
\author{Peter A. Ivanov}
\affiliation{Department of Physics, St. Kliment Ohridski University of Sofia, James Bourchier 5 blvd, 1164 Sofia, Bulgaria}

\begin{abstract}
I propose a quantum sensor based on driven-dissipative quantum system for the joint estimation of two conjugated variables characterizing the phase space displacement. The quantum probe consists of lattice system with two level atoms and bosonic modes which interact via dipolar coupling. Interplay between the coherent dynamics and dissipative processes of losses of bosonic excitations leads to a steady state which exhibits a non-analytical behaviour. I show that close to the dissipative phase transition the sensitivity of one of the conjugated parameters either the magnitude of the phase of the displacement can be significantly enhanced. Moreover, I show that the sum of the measurement uncertainties of the two parameters can overcome the standard quantum limit.
\end{abstract}

%\pacs{
%03.67.Ac, %Quantum computation architectures and implementations
%03.67.Bg,
%03.67.Lx,
%42.50.Dv %Coherent control of atomic interactions with photons
%}
\maketitle

%%%%%%%%%%%%%%%%%%%%%%%%%%%%%%%%%%%%%%%%%%%%%%%%%%%%%%%%%%%%%%%%%%%%%%%%%%%
%%%%%%%%%%%%%%%%%%%%%%%%%%%%%%%%%%%%%%%%%%%%%%%%%%%%%%%%%%%%%%%%%%%%%%%%%%%
%%%%%%%%%%%%%%%%%%%%%%%%%%%%%%%%%%%%%%%%%%%%%%%%%%%%%%%%%%%%%%%%%%%%%%%%%%%
%========================================================================
%========================================================================
\section{Introduction}
Quantum sensing is one of the most promising application of quantum technologies. Usually quantum metrology task involves estimation of a single parameter. High-precision quantum estimation can be achieved by exploiting quantum critical systems which exhibit quantum phase transition as a probe. Indeed, as was shown in \cite{Zanardi2008,Ivanov2013,Macieszczak2016,Garbe2020} the sensitivity of single parameter estimation can be significantly improved close to a quantum critical point. However, in general physical process can involve the simultaneous estimation of multiple parameters, which gives rise to the emergent field of multiparameter quantum metrology. A large class of quantum metrology problems involve joint estimation of more than one parameter, including for example enhanced estimation of multiple phases \cite{Humphreys,Gessner2018,Pezze2017}, phases and noises \cite{Vidrighin2014,Yue2014,Genoni2011}, multidimensional field \cite{Baumgratz2016} as well as the estimation of the phase space displacement parameters \cite{Genoni2013,Bradshaw2018,Ivanov2018} (see the recent reviews on multiparameter quantum metrology \cite{Szczykulska2016,Albarelli2020}). In analogous with enhanced single parameter estimation a natural task arises to quantify the sensitivity of the multiparameter estimation close to a quantum critical point.

In this work I discuss the estimation of two conjugated parameters characterizing the phase space displacement using quantum probe which exhibit dissipative phase transition. Such a new class of phase transitions emerges due to the interplay between the coupling with the environment and the driving mechanics in an open quantum systems. The dissipative phase transitions are characterized with a non-analytical change in the steady state \cite{Minganti2018} and can be used as a potential resource for high precision quantum metrology. Our dissipative quantum probe consists of one dimensional lattice system where at each site a single two level atom interact via dipolar coupling with a bosonic mode. The coupling between the bosons at different lattice sites is provided via hopping processes. The interplay between the coherent dynamics and dissipative processes which causes losses of bosonic excitations leads to non-equilibrium regime where the information of the two parameters is encoded in the \emph{steady state} density matrix elements. I consider the limit in which the spin excitations are highly suppressed such that the system approaches bosonic multimode Gaussian steady state. Crucially, the effect of the phase space displacement is to break explicitly the parity symmetry of the lattice model which leads to a non vanishing expectation values of the quadratures. First I discuss a single mode case where critical point separates normal to superradiant dissipative phase transition \cite{Hwang2015,Hwang2018}. I show that close to the critical coupling the average quadratures are enhanced and essentially diverge approaching the dissipative phase transition, which can be used to improve the sensitivity in the single parameter displacement estimation  \cite{Ivanov2020}. In order to quantify the uncertainty of the two-parameter estimation I use quantum Fisher information matrix (QFIM) which can be explicitly derived. I show that thanks to the spin-boson coupling the uncertainty of the joint estimation can be improved compared to the non-driven case. Moreover, for coupling closed to the critical point the sensitivity of one of the conjugated parameters either the magnitude or the phase of the displacement can be significantly enhanced. As a result of that our two-parameter estimation technique can operate beyond the quantum standard limit.

Furthermore, I extend the quantum sensing technique by including the hopping between the bosons at different lattice sites. Approaching the steady-state the system is described by the multimode Gaussian state. I consider the two coupled lattice sites and show that the covariant matrix is independent on the parameters we wish to estimate which significantly simplifies the expression for the QFIM. All elements of the QFIM diverge for spin-boson coupling approaching the critical point signals the occurrence of dissipative phase transition. I show that the critical point is modified by the hopping and its value can be lowered compared to the single mode case. Moreover, I show that the sensitivity of the two parameter estimation can be improved compared to the achievable ultimate precision using two uncoupled quantum probes.

The parer is organized as follows: In Sec. \ref{framework} I provide the general theoretical framework for multiparameter quantum estimation. In Sec. \ref{QSP} is presented the quantum probe consisting of coupled light-matter system which exhibits dissipative phase transition. In Sections \ref{SM} and \ref{MM} I discuss the sensitivity of the two-parameter estimation in terms of QFIM. I show that close to the dissipative phase transition one can achieve significant enhancement of the sensitivity of one of the conjugated parameters. It is shown that thanks to the driven-dissipative dynamics the sum of the measurement uncertainties can overcome the standard quantum limit. Finally, the conclusions are presented in Sec. \ref{C}.

\section{Generalized Theoretical Framework for Multiparameter Quantum Estimation}\label{framework}
In order to perform quantum multiparameter estimation of $p$ unknown parameters $\bold{q}=(q_{1},q_{2},\ldots,q_{p})$ one needs a quantum probe described by a density matrix $\hat{\rho}_{0}$. Upon the action of the time evolution the quantum probe evolves into the state $\hat{\rho}_{q}$ where the information of the parameters are encoded in the density matrix elements.
The sensitivity of the estimator is described by the covariance matrix which elements are ${\rm Var}(\bold{q})_{ij}=\langle q_{i}q_{j}\rangle-\langle q_{i}\rangle\langle q_{j}\rangle$ where the diagonal elements quantifies the uncertainty of the estimation of the individual parameters while the off-diagonal elements indicates a possible correlation between the different parameters. The ultimate precision in the multiparameter estimation is quantified by the Cramer-Rao bound \cite{Szczykulska2016,Albarelli2020}
\begin{equation}
{\rm Var}(\bold{q})\geq (\nu \mathcal{F} )^{-1},
\end{equation}
where $\nu$ is the number of experimental repetitions and $\mathcal{F}_{ij}$ is the $p\times p$ real-valued symmetric QFIM.

In order to calculate the multiparameter QFIM we define the hermitian symmetric logarithmic derivative (SLD) operator $\hat{\mathcal{L}}_{q_{k}}$ for each of the parameters which obeys the operator equation \cite{Paris2009}
\begin{equation}
2\frac{\partial\hat{\rho}_{q}}{\partial q_{k}}=\hat{\mathcal{L}}_{q_{k}}\hat{\rho}_{q}+\hat{\rho}_{q}\hat{\mathcal{L}}_{q_{k}}.
\end{equation}
For the density matrix with spectral decomposition $\hat{\rho}_{q}=\sum_{n}p_{n}|\psi_{n}\rangle\langle\psi_{n}|$ the SLD can be expressed as
\begin{equation}
\hat{\mathcal{L}}_{q_{k}}=2\sum_{n,m}\frac{\langle\psi_{n}|\partial_{q_{k}}\hat{\rho}_{q}|\psi_{m}\rangle}{p_{n}+p_{m}}|\psi_{n}\rangle\langle\psi_{m}|.\label{SLD}
\end{equation}
Then using the SLD operators one can express the real and symmetric QFIM elements as follows
\begin{equation}
\mathcal{F}_{km}=\frac{1}{2}{\rm Tr}(\hat{\rho}_{q}\{\hat{\mathcal{L}}_{q_{k}},\hat{\mathcal{L}}_{q_{m}}\}),
\end{equation}
where $\{\cdot,\cdot\}$ is the anticommutator. We point out that QFIM can be interpreted as a measure of distinguishability of two quantum states with respect of infinitesimal change of the parameters of interest. Indeed, one can define Bures distance between two infinitesimally close quantum states by $ds^{2}_{\rm B}=\sum_{k,m}g_{km}dq_{k}dq_{m}$ where $g_{km}=\frac{1}{4}\mathcal{F}_{km}$ is the metric tensor \cite{Braunstein1994}. This intimate relation between distance and QFIM indicates that the quantum parameter estimation can be substantially enhanced close to phase transition where infinitesimally small change of parameters give rise to huge change of the distance \cite{Zanardi2008}.

For single parameter estimation the optimal measurement is always achieved by the projective measurements composed by the eigenvectors of SLD operator. However, for multiparameter estimation the SLD operators corresponding to different physical observable may not commute and hence the ultimate precision is achieved by incompatible measurements. This is hold for conjugated variables for which a Heisenberg-type uncertainty relation applies. Defining
\begin{equation}
\hat{\mathcal{C}}_{q_{k},q_{m}}=[\hat{\mathcal{L}}_{q_{k}},\hat{\mathcal{L}}_{q_{m}}],
\end{equation}
sufficient condition to exist an optimal measurement which saturates the quantum Cramer-Rao bound is the commutativity of all pairs of the SLD operators, $\hat{\mathcal{C}}_{q_{k},q_{m}}=0$. A weak condition for the saturation of the multiparameter quantum Cramer-Rao bound requires the commutativity of the SLD operators on average, ${\rm Tr}(\hat{\rho}_{q}\hat{\mathcal{C}}_{q_{k},q_{m}})=0$ \cite{Szczykulska2016,Matsumoto2002,Ragy2016}.

In the following I will discuss two-parameter estimation of the magnitude and the phase of unknown displacement using open quantum system as a probe, which exhibits dissipative phase transition. The quantum probe consists of chain of dissipative coupled light-matter systems each of them described by the quantum Rabi model.
\section{Quantum Sensing Protocol}\label{QSP}
Consider a linear chain of $N$ spins each coupled with a single bosonic mode via dipolar interaction described by quantum Rabi model. The bosons at different lattice sites are coupled due to the hopping processes subject to the tight-binding model. The total Hamiltonian then is given by
\begin{eqnarray}
\hat{H}&=&\hbar\sum_{k=1}^{N}\{\omega \hat{a}_{k}^{\dag}\hat{a}_{k}+\frac{\Omega}{2} \sigma_{k}^{z}+g(\hat{a}^{\dag}_{k}+\hat{a}_{k})\sigma_{k}^{x}\notag\\
&&+\frac{F}{2}(\hat{a}^{\dag}_{k}e^{i\chi}+\hat{a}_{k}e^{-i \chi})\}+\hbar\sum_{k>l}^{N}\kappa_{kl}(\hat{a}^{\dag}_{k}\hat{a}_{l}+\hat{a}^{\dag}_{k}\hat{a}_{l}),\label{Rabi}
\end{eqnarray}
where $\omega$ is the frequency of the bosonic field, $\hat{a}^{\dag}_{k}$ and $\hat{a}_{k}$ are creation and annihilation operators of the bosonic excitation at the $k$th site, $\sigma^{x,z}_{k}$ are the Pauli matrices associates with the $k$th spin and $g$ is the coupling strength. $\Omega$ is the frequency of the spin and $\kappa_{kl}$ is the hopping strength. The two conjugated parameters which we wish to estimate are the magnitude of the displacement $q_{1}=F$ and respectively its phase $q_{2}=\chi$. Such a displacement term can be created for example by applying a time-varying force with unknown magnitude and phase which displaces motion amplitude of the quantum oscillators \cite{Maiwald2009,Ivanov2015,Ivanov2016,Wolf2019,Burd2019}.
The effect of decay of bosonic excitation is described within the framework of master equation in Lindblad form,
\begin{equation}
\partial_{t}\hat{\rho}_{q}=-\frac{i}{\hbar}[\hat{H},\hat{\rho}_{q}]+\sum_{k=1}^{N}\gamma_{k}\hat{\mathcal{L}}[\hat{a}_{k}]\hat{\rho}_{q},\label{lindbad}
\end{equation}
where the Lindblad term for each bosonic mode is $\hat{\mathcal{L}}[\hat{a}_{k}]\hat{\rho}_{q}=2\hat{a}_{k}\hat{\rho}_{q}\hat{a}^{\dag}_{k}-\{\hat{a}^{\dag}_{k}\hat{a}_{k},\hat{\rho}_{q}\}_{+}$ and $\gamma_{k}$ is the decay rate.

Such a driven-dissipative quantum probe can be naturally implemented with various quantum optical systems. For example one possible experimental setup relies on using laser cooled trapped ions. In that case the bosonic degree of freedom is provided by the local phonons with Coulomb mediated phonon hopping \cite{Ivanov2009,Haze2012} and the spins are implemented by the internal levels of the trapped ions. Engineering of the dissipation can be implemented via sympathetic cooling of the ion's oscillations which introduces motion damping \cite{Lemmer2015}. Other suitable system for the realization of the quantum probe is dissipative cavity and circuit QED systems where the bosons are represented by the quantized modes, while the spins are implemented by real two-level atoms, or artificial atoms such as quantum dots or superconducting circuits \cite{Noh2017}.

In our quantum metrology scheme the system is prepared initially in state with density matrix $\hat{\rho}(0)=\hat{\rho}_{\rm spin}\otimes\hat{\rho}_{\rm b}$ and then evolves according to the master equation (\ref{lindbad}). Here $\hat{\rho}_{\rm spin}=\otimes_{k=1}^{N}\left|\downarrow_{k}\right\rangle\left\langle\downarrow_{k}\right|$, where $\sigma^{z}_{k}\left|\downarrow_{k}\right\rangle=-\left|\downarrow_{k}\right\rangle$ and respectively $\hat{\rho}_{\rm b}=\otimes_{k=1}^{N}\left|0_{k}\right\rangle\left\langle0_{k}\right|$ with $\left|n_{k}\right\rangle$ being the Fock state for the $k$th boson. Defining the dimensionless coupling $\lambda=2g/\sqrt{\omega\Omega}$ and consider the limit $\eta=\omega/\Omega\rightarrow 0$ one can trace out the spin degree of freedom. Indeed, making the unitary transformation $\hat{U}=\prod_{k}^{N} e^{i\frac{g}{\Omega}(\hat{a}^{\dag}_{k}+\hat{a}_{k})\sigma_{k}^{y}}$ such that the effective Hamiltonian $\hat{H}_{\rm eff}=\hat{U}\hat{H}\hat{U}^{-1}$ becomes
\begin{eqnarray}
\hat{H}_{\rm eff}&=&\hbar\sum_{k=1}^{N}\{\tilde{\omega}\hat{a}^{\dag}_{k}\hat{a}_{k}-\frac{\omega\lambda^{2}}{4}(\hat{a}^{\dag2}_{k}+\hat{a}^{2}_{k})
+\frac{F}{2}(\hat{a}^{\dag}_{k}e^{i\chi}+\hat{a}_{k}e^{-i\chi})\}\notag\\
&&+\hbar\sum_{k>l}^{N}\kappa_{kl}(\hat{a}^{\dag}_{k}\hat{a}_{l}+\hat{a}^{\dag}_{k}\hat{a}_{l}),\label{Heff}
\end{eqnarray}
where $\tilde{\omega}=\omega(1-\lambda^{2}/2)$.
\begin{figure}
\includegraphics[width=0.45\textwidth]{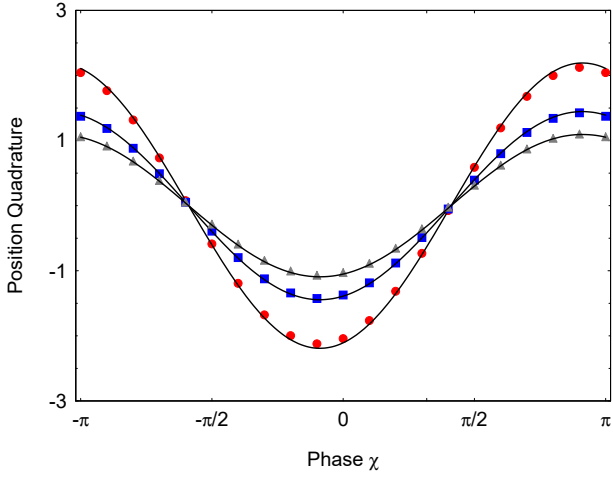}
\caption{(Color online) Steady-state position quadrature versus the phase $\chi$. We compare the exact solution derived from the original Hamiltonian (\ref{Rabi}) with $\lambda=0.85$ (grey triangle), $\lambda=0.9$ (blue squares), $\lambda=0.95$ (red circles) and the steady-state solution $\langle \hat{x}\rangle=2\alpha\cos(\delta)$ (solid line). The other parameters are set to $\lambda_{\rm c}=1.04$ and $\tilde{F}=0.38$.}
\label{fig1}
\end{figure}

Interplay between the coherent and dissipative dynamics leads to a non-equilibrium steady-state of the system which can exhibits a non analytical behaviour. Our quantum sensing protocol relies on the time evolution of the system into the steady state where the two-parameter estimation is performed. Since the dynamics is quadratic in the bosonic operators the steady state of the system is of Gaussian form and the density operator can be reconstructed from the first and the second moments. In order to describes the $N$-mode Gaussian state of the system we define quadrature operator $\hat{R}=\{\hat{x}_{1},\hat{p}_{1},\ldots,\hat{x}_{N},\hat{p}_{N}\}^{\rm T}$ and mean displacement vector $\bold{d}=\langle \hat{\bold{R}}\rangle$ \cite{Weedbrook2012}. Here $\hat{x}_{k}=(\hat{a}^{\dag}_{k}+\hat{a}_{k})$ and $\hat{p}_{k}=i(\hat{a}^{\dag}_{k}-\hat{a}_{k})$ are the position and momentum quadratures for $k$th oscillator. Then the covariant matrix becomes
\begin{equation}
V_{kl}=\frac{1}{2}\langle \hat{R}_{k}\hat{R}_{l}+\hat{R}_{l}\hat{R}_{k}\rangle-d_{k}d_{l}.\label{covariant}
\end{equation}
Finally, using the covariant matrix and mean displacement one can quantified the sensitivity of the two-parameter estimation in terms of QFIM.
\section{Single Mode Case}\label{SM}
We begin by consider the non-equilibrium steady state of the system for $\kappa_{kl}=0$. In that case the single mode Gaussian steady-state can be expressed as $\hat{\rho}_{q}=\hat{R}(\delta)\hat{D}(\alpha)\hat{S}(\xi)\hat{\nu} \hat{S}^{\dag}(\xi) \hat{D}^{\dag}(\alpha) \hat{R}^{\dag}(\delta)$, where $\hat{R}(\delta)=e^{i\delta \hat{a}^{\dag}\hat{a}}$ is the rotation operator, $\hat{D}(\alpha)=e^{\alpha(\hat{a}^{\dag}-\hat{a})}$ is the displacement operator, $\hat{S}(\xi)=e^{\frac{r}{2}(\hat{a}^{2}e^{-2i\phi}-\hat{a}^{\dag2}e^{2i\phi})}$ is the single mode squeezing operator and $\hat{\nu}=\sum_{n}p_{n}|n\rangle\langle n|$ is the thermal state. Here $p_{n}=N_{\rm th}^{n}/(1+N_{\rm th})^{n+1}$ is the thermal state probability and $N_{\rm th}$ stand for the average number of thermal excitations, (see Appendix \ref{SLD_App} for details). Note that $N_{\rm th}$ is independent on the parameters we wish to estimate. I find that the displacement amplitude and rotation phase angle are given by
\begin{eqnarray}
&&\alpha=\frac{\tilde{F}}{2(\lambda^{2}_{\rm c}-\lambda^{2})}\sqrt{\lambda^{2}_{\rm c}-\lambda^{2}\tilde{\gamma}\sin(2\chi)+\lambda^{2}(\lambda^{2}-2)\sin^{2}(\chi)},\notag\\
&&\tan(\delta)=\frac{(\lambda^{2}-1)\sin(\chi)-\tilde{\gamma}\cos(\chi)}{\tilde{\gamma}\sin(\chi)-\cos(\chi)}
\end{eqnarray}
and respectively the squeezing and its phase are
\begin{eqnarray}
\tanh(2r)=\frac{\lambda^{2}}{\sqrt{4(\lambda_{\rm c}^{2}-\lambda^{2})+\lambda^{4}}},\quad\tan(2\phi+2\delta)=\frac{2\tilde{\gamma}}{2-\lambda^{2}}\label{squeezing},
\end{eqnarray}
with $\tilde{F}=F/\omega$, $\tilde{\gamma}=\gamma/\omega$. Here $\lambda^{2}_{\rm c}=1+\tilde{\gamma}^{2}$ is the critical coupling which separates normal $\lambda\leq\lambda_{\rm c}$ to superradiant $\lambda>\lambda_{\rm c}$ dissipative phase transition. Note that the present estimation scheme is focus on the case $\lambda\leq\lambda_{\rm c}$. We emphasize that the information of the two parameters we wish to estimate is encoded in three parameters $\delta$, $\alpha$ and $\phi$ which is in contrast with the standard two parameters phase space estimation where the parameters are encoded respectively in the amplitude and the phase of the displacement \cite{Genoni2013,Ivanov2018}. In Fig. \ref{fig1} is shown comparison between the exact and analytical results for the steady-state position quadrature for different phase $\chi$. We see that by increasing $\lambda$ the displacement amplitude is enhanced and respectively diverges approaching the dissipative phase transition at the critical coupling $\lambda_{\rm c}$.

Having the expression for the non-equilibrium steady state one can derive the expressions for the corresponding two SLD operators. Indeed, using (\ref{SLD}) it is straightforward to prove that
\begin{equation}
\hat{\mathcal{L}}_{F}=\frac{2\partial_{F}\alpha}{1+2N_{\rm th}}\hat{R}(\delta)\hat{D}(\alpha)\hat{S}(\xi)(\beta \hat{a}^{\dag}+\beta^{*}\hat{a})\hat{S}^{\dag}(\xi)\hat{D}^{\dag}(\alpha)\hat{R}^{\dag}(\delta),\label{SLD_F}
\end{equation}
with $\beta(r,\phi)=\cosh(r)+e^{2i\phi}\sinh(r)$ and respectively
\begin{equation}
\hat{\mathcal{L}}_{\chi}=\frac{2}{1+2N_{\rm th}}\hat{R}(\delta)\hat{D}(\alpha)\hat{S}(\xi)(\upsilon \hat{a}^{\dag}+\upsilon^{*}\hat{a})\hat{S}^{\dag}(\xi)\hat{D}^{\dag}(\alpha)\hat{R}^{\dag}(\delta),\label{SLD_chi}
\end{equation}
with $\upsilon=(\partial_{\chi}\alpha)\beta(r,\phi)+i\alpha(\partial_{\chi}\delta)\beta(-r,\phi)$, (see Appendix \ref{SLD_App} for an overview of the derivation). In order to quantify the sensitivity of the two-parameter estimation we define the inverse QFIM
\begin{equation}
(\mathcal{F}^{-1})_{km}=\left(
\begin{array}{cc}
\mathcal{F}^{-1}_{F} & \mathcal{F}^{-1}_{F\chi} \\
\mathcal{F}^{-1}_{\chi F} & \mathcal{F}^{-1}_{\chi}
\end{array}%
\right).\label{QFImatrix}
\end{equation}
\begin{figure}
\includegraphics[width=0.45\textwidth]{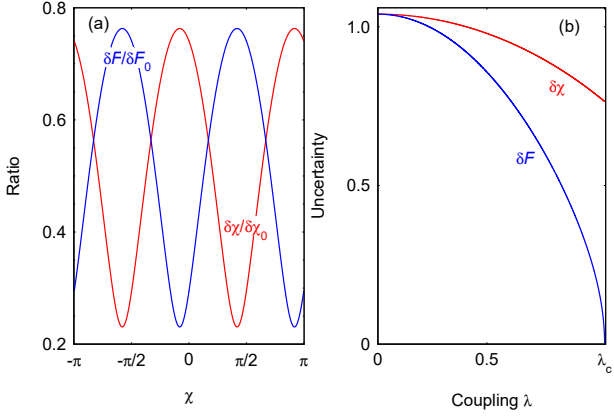}
\caption{(Color online) (a) Ratio $\delta F/\delta F_{0}$ and $\delta\chi/\delta\chi_{0}$ as a function of phase $\chi$. The parameters are set to $\lambda_{\rm c}=1.04$ and $\lambda=0.95\lambda_{\rm c}$. (b) Minimal detectable parameters $\delta F$ in units of $\omega$ and $\delta\chi$ according Eq. (\ref{sensitivty}) as a function of $\lambda$. The force sensitivity is improved for phase $\chi$ close to the optimal given by $\tan(2\chi_{\rm opt})=2\tilde{\gamma}/(\lambda^{2}_{\rm c}-2)$. }
\label{fig2}
\end{figure}The diagonal elements of the inverse QFIM provides achievable bounds for the sensitivity of the joint estimation. We find (see Appendix \ref{SLD_Operators} for details)
\begin{eqnarray}
&&\mathcal{F}^{-1}_{F}=\frac{\omega^{2}}{4}\{4(\lambda^{2}_{\rm c}-\lambda^{2})+\lambda^{4}+\lambda^{2}((\lambda^{2}-2)\cos(2\chi)\notag\\
&&\quad\quad+2\tilde{\gamma}\sin(2\chi))\},\notag\\
&&\mathcal{F}^{-1}_{\chi}=\frac{\omega^{2}}{4F^{2}}\{4(\lambda^{2}_{\rm c}-\lambda^{2})+\lambda^{4}-\lambda^{2}((\lambda^{2}-2)\cos(2\chi)\notag\\
&&\quad\quad+2\tilde{\gamma}\sin(2\chi))\},\label{sensitivty}
\end{eqnarray}
Hence the ultimate achievable precision becomes $\delta F^{2}\ge\frac{1}{\nu}\mathcal{F}^{-1}_{F}$ and $\delta \chi^{2}\ge\frac{1}{\nu}\mathcal{F}^{-1}_{\chi}$. The off-diagonal elements of the matrix (\ref{QFImatrix}) describe the correlation between the two parameters. We obtain
\begin{equation}
\mathcal{F}^{-1}_{F\chi}=-\frac{\omega^{2}\lambda^{2}}{4F}\{(\lambda^{2}-2)\sin(2\chi)-2\tilde{\gamma}\cos(2\chi)\}.
\end{equation}

As a comparison I first discuss the ultimate precision by setting $g=0$ and thus $\lambda=0$. This correspond to a quantum probe consisting of a single dissipative harmonic oscillator sensitive to the magnitude and the phase of unknown displacement. Using Eq. (\ref{sensitivty}) one can show that the uncertainty of the parameters estimation is given by $\delta F_{0}\ge\frac{\omega}{\sqrt{\nu}}\lambda_{\rm c}$ and $\delta \chi_{0}\ge\frac{\omega}{\sqrt{\nu} F}\lambda_{\rm c}$. Crucially enhancement of the joint sensitivity can be realized by increasing coupling $\lambda$. In Fig. \ref{fig2}(a) is shown the ratio $\delta F/\delta F_{0}$ and $\delta \chi/\delta \chi_{0}$ according to Eq. (\ref{sensitivty}). We see that the ultimate joint sensitivity which is achieved by the driven dissipative dynamics can be improved compared to the non-driven case with $\delta F_{0}$ and $\delta \chi_{0}$. In particular, when the phase $\chi$ is closed to the optimal phase given by $\tan(2\chi_{\rm opt})=2\tilde{\gamma}/(\lambda^{2}_{\rm c}-2)$ one can achieve significantly improve sensitivity of one of the parameters. Indeed, close to the critical coupling $\lambda_{\rm c}$ the two parameters correlations vanishes $\mathcal{F}^{-1}_{F\chi}\approx 0$ and the uncertainty of the joint estimation of the displacement magnitude and the phase becomes
\begin{eqnarray}
&&\delta F^{2}\ge\frac{\omega^{2}}{4}\{4(\lambda^{2}_{\rm c}-\lambda^{2})+\lambda^{4}-\lambda^{2}\sqrt{4(\lambda^{2}_{\rm c}-\lambda^{2})+\lambda^{4}}\},\notag\\
&&\delta \chi^{2}\ge\frac{\omega^{2}}{4F^{2}}\{4(\lambda^{2}_{\rm c}-\lambda^{2})+\lambda^{4}+\lambda^{2}\sqrt{4(\lambda^{2}_{\rm c}-\lambda^{2})+\lambda^{4}}\}.\label{uncertainty}
\end{eqnarray}
Approaching the dissipative phase transition the joint sensitivity scales according to $\delta F\sim \omega\sqrt{\lambda_{\rm c}-\lambda}$ and $\delta\chi\sim \frac{\omega}{\sqrt{2}F}\lambda_{\rm c}^{2}$. Hence in this limit the quantum probe becomes sensitive to infinitely small force perturbation, see Fig. \ref{fig2}(b). Also we observe that as long as $\gamma<\omega$ we have $\delta\chi<\delta\chi_{0}$ such that the phase sensitivity is improved compared to $\delta\chi_{0}$. Note that for phase equal to $\chi=\chi_{\rm opt}+\pi/2$ one can show that $\delta F\sim \frac{\omega}{\sqrt{2}}\lambda_{\rm c}^{2}$ and $\delta\chi\sim \frac{\omega}{F}\sqrt{\lambda_{\rm c}-\lambda}$ and thus one can enhance respectively the phase sensitivity.

Furthermore, we evaluate the commutator of the SLD operators corresponding to the two displacement parameters. Using Eqs. (\ref{SLD_F}) and (\ref{SLD_chi}) we obtain
\begin{equation}
\hat{\mathcal{C}}_{F\chi}=\frac{8i F}{\omega^{2}}\frac{\hat{\bold{1}}}{4(\lambda^{2}_{\rm c}-\lambda^{2})+\lambda^{4}}. \label{comm}
\end{equation}
Since we deal with conjugate variables for which a Heisenberg uncertainty relation holds the two SLD operators do not commute even in an average. Using (\ref{comm}) one can estimate the commutator close to the dissipative phase transition, $\lambda\rightarrow \lambda_{\rm c}$. We find
\begin{equation}
\hat{\mathcal{C}}_{F\chi}\sim \frac{8iF\omega^{2}}{(\omega^{2}+\gamma^{2})^{2}}\hat{\bold{1}}.
\end{equation}
We note in order to satisfy the condition of weak commutativity $\langle\hat{\mathcal{C}}_{F\chi}\rangle=0$ one can lower $\omega$ which on one hand will improve the sensitivity of one of the parameters, for example $\delta F$ but on the other hand will spoil the phase estimation because $\delta \chi\sim 1/\omega$.

Finally, one can evaluate the sum of the measurement uncertainties of the two parameters. For this goal it is convenient to introduce dimensionless quantities $q=\tilde{F}\cos(\chi)$ and $p=\tilde{F}\sin(\chi)$. Then it is straightforward to show that the QFIM elements do not dependent on the values of the two parameters to be estimated. We find that both uncertainties becomes
\begin{equation}
\delta q^{2}\geq\frac{1}{2\nu}(2\lambda^{2}_{\rm c}-3\lambda^{2}+\lambda^{4}),\quad \delta p^{2}\geq\frac{1}{2\nu}(2\lambda^{2}_{\rm c}-\lambda^{2})
\end{equation}
and therefore
\begin{equation}
\delta q^{2}+\delta p^{2}\geq\frac{1}{2\nu}\{4(\lambda^{2}_{\rm c}-\lambda^{2})+\lambda^{4}\}.
\end{equation}
The quantum standard limit (SQL) requires $\delta q^{2}+\delta p^{2}\geq\frac{2}{\nu}$ \cite{Genoni2013}. The latter has simple explanation, namely it corresponds to the ultimate achievable precision for non-driven quantum probe with $\lambda=0$ and $\lambda_{\rm c}=1$. Crucially the effect of the spin-boson coupling $\lambda$ is to improve simultaneously the uncertainty of the two parameters displacement estimation. Indeed, approaching the dissipative phase transition the sum of the variances becomes $\delta q^{2}+\delta p^{2}\sim\frac{\lambda^{4}_{\rm c}}{2\nu}$ and thus as long as $\gamma<\omega$ one can overcome the SQL. Note that the beating of the SQL is equivalent that both uncertainties on the estimation of the parameters $q$ and $p$ are $\delta q<1$ and $\delta p<1$.
\begin{figure}
\includegraphics[width=0.45\textwidth]{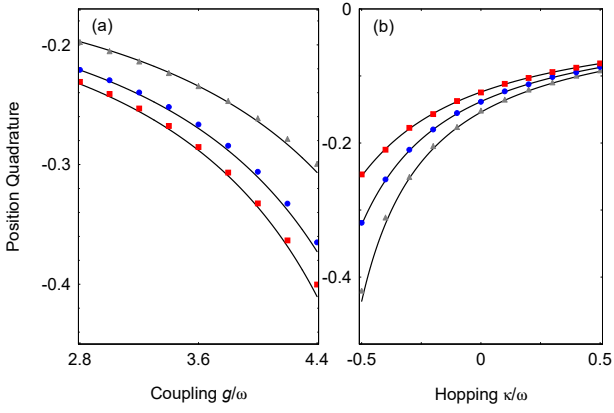}
\caption{(Color online) (a) Average position quadrature as a function of the coupling $g/\omega$ for different hopping amplitude $\kappa$. We compare the exact solution of the master equation with Hamiltonian (\ref{Rabi}) with the steady-state result (\ref{x_kappa}) (solid lines). The parameters are set to $\kappa/\omega=-0.4$ (red squares), $\kappa/\omega=-0.47$ (blue circles), and $\kappa/\omega=-0.5$ (grey triangles). The other parameters are $\eta=4\times 10^{-3}$, $\tilde{F}=0.13$, $\tilde{\gamma}=0.16$ and $\chi=\pi/7$. (b) The same but now vary the hopping amplitude for different couplings, $g/\omega=3.1$ (black triangles), $g/\omega=3.9$ (blue circles), $g/\omega=4.5$ (red squares).}
\label{fig3}
\end{figure}

\section{Multi-Mode Case}\label{MM}
Let's extend the two-parameter estimation by considering coupled system described with Hamiltonian (\ref{Rabi}). Again in the limit $\eta\rightarrow 0$ the model is transformed into the dissipative system of harmonic oscillators which interact via hopping dynamics. Hereafter we assume nearest neighbour hopping between the sites, namely $\kappa_{kl}=\kappa\delta_{k,l+1}$. Note that depending on the physical realization of the scheme the sign of the hopping can vary. Indeed, for quantum probe based on trapped ion system the hopping can be positive (negative) depending on the either we use radial (axial) phonons as a bosonic degree of freedom. For realization with coupled cavity array the sigh of the hopping is negative.
\begin{figure}
\includegraphics[width=0.45\textwidth]{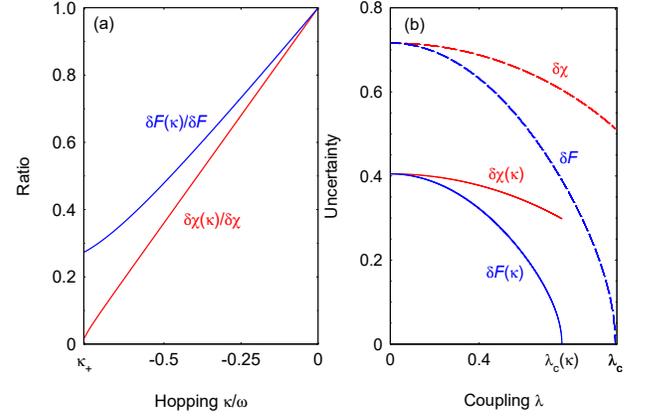}
\caption{(Color online) (a) Ratios $\delta F(\kappa)/\delta F$ and $\delta \chi(\kappa)/\delta \chi$ as a function of the hopping according Eq. (\ref{un_kappa}). The parameters are set to $\tilde{\gamma}=0.16$, $\lambda=0.59$ and $\chi=\pi/3$. (b) Minimal detectable force and phase versus the coupling $\lambda$ (solid lines) for $\tilde{\kappa}=-0.45$. As a comparison is shown the results for $\kappa=0$ (dashed lines).}
\label{fig4}
\end{figure}

Using Eq. (\ref{Heff}) one can derive the set of equations for the expectation values of the bosonic operators. We have
\begin{eqnarray}
&&\partial_{t}\langle \hat{a}^{\dag}_{s}\rangle=\{i\left(1-\frac{\lambda^{2}}{2}\right)-\gamma_{s}\}\langle \hat{a}^{\dag}_{s}\rangle-i\frac{\omega\lambda^{2}}{2}\langle\hat{a}_{s}\rangle+i\kappa\langle\hat{a}^{\dag}_{2}\rangle\notag\\
&&\quad\quad\times\delta_{2,s+1}+i\kappa\langle\hat{a}^{\dag}_{N-1}\rangle\delta_{N-1,s-1}+i\frac{F}{2}e^{-i\chi},\quad s=1,N,\notag\\
&&\partial_{t}\langle \hat{a}^{\dag}_{k}\rangle=\{i\left(1-\frac{\lambda^{2}}{2}\right)-\gamma_{k}\}\langle \hat{a}^{\dag}_{k}\rangle-i\frac{\omega\lambda^{2}}{2}\langle\hat{a}_{k}\rangle+i\kappa\langle\hat{a}^{\dag}_{k-1}\rangle\notag\\
&&\quad\quad+i\kappa\langle\hat{a}^{\dag}_{k+1}\rangle+i\frac{F}{2}e^{-i\chi},\quad k\neq 1, N
\end{eqnarray}
In the steady state the information of the two parameters is encoded in the multimode Gaussian state. Although the steady-state quadratures can be found for any number of sites the respective expressions are too complicated to be presented here. As an example consider two lattice site where the steady state position quadrature is given by
\begin{eqnarray}
\langle \hat{x}_{k}\rangle=-\tilde{F}\frac{(1+\tilde{\kappa})\cos(\chi)-\tilde{\gamma}\sin(\chi)}{(1+\tilde{\kappa})(\lambda^{2}_{+}(\kappa)-\lambda^{2})},\label{x_kappa}
\end{eqnarray}
with $\tilde{\kappa}=\kappa/\omega$ and for simplicity we set $\gamma_{k}=\gamma$ ($k=1,2$). Here $\lambda_{\rm +}(\kappa)=\sqrt{\tilde{\gamma}^{2}+(1+\tilde{\kappa})^{2}}/\sqrt{1+\tilde{\kappa}}$ is the critical coupling which is modified by the hopping $\kappa$ compared to the single mode case with $\lambda_{\rm c}$. As long as $\tilde{\kappa}>-1$, the critical coupling $\lambda_{+}(\kappa)$ is real such that the quadratures diverge in the limit $\lambda\rightarrow\lambda_{+}(\kappa)$ signal the existence of dissipative phase transition. For $\tilde{\kappa}<-1$ the coupling $\lambda_{+}(\kappa)$ becomes purely imaginary and thus no enhancement of the average quadratures is possible by increasing $\lambda$. We observe that as long as $\tilde{\kappa}<0$ and $\tilde{\kappa}>\tilde{\kappa}_{\rm min}$ where $\tilde{\kappa}_{\rm min}=-1+\tilde{\gamma}^{2}$, the critical coupling is smaller compared to $\lambda_{\rm c}$, namely $\lambda_{+}(\kappa)<\lambda_{\rm c}$. In Fig. \ref{fig3}(a) is shown comparison between the exact result with original Hamiltonian (\ref{Rabi}) and steady state position quadrature (\ref{x_kappa}) as a function of $g/\omega$ for different $\kappa$. We see that for $\eta\ll 1$ the exact dynamics is described very closely with the effective Hamiltonian (\ref{Heff}). We observe that by increasing $|\kappa|$ the critical coupling $\lambda_{+}$ decreases which leads to higher average position quadrature $|\langle \hat{x}_{k}\rangle|$. As we will see below smaller value of $\lambda_{+}(\kappa)$ can leads to better sensitivity in a sense that for the same value of $\lambda$ the two parameters displacement estimation is improved compared to the single mode case.

Furthermore, it is straightforward to show that the denominator in Eq. (\ref{x_kappa}) can be rewritten as $(1+\tilde{\kappa})(\lambda_{+}^{2}(\kappa)-\lambda^{2})=(\kappa_{+}-\tilde{\kappa})(\kappa_{-}-\tilde{\kappa})$. Here we define critical hopping amplitudes $\kappa_{\pm}=\frac{1}{2}\{\lambda^{2}-2\pm\sqrt{\lambda^{4}-4\tilde{\gamma}^{2}}\}$ which are reals as long as $\lambda^{4}\ge 4\tilde{\gamma}^{2}$.
As is shown in Fig. \ref{fig3}(b) increasing the hopping amplitude $\kappa$ the average position quadrature $|\langle \hat{x}_{k}\rangle|$ increases and eventually diverges in the limit $\tilde{\kappa}\rightarrow\kappa_{+}$. Note that in Fig. \ref{fig3}(b) the parameters are set such that $\kappa_{+}>\kappa_{-}$ and $\kappa_{\pm}<0$.

In order to describe the sensitivity of the two-parameter estimation in terms of QFIM one need to find the covariant matrix (\ref{covariant}) for the two mode Gaussian state. I find that all elements of $V_{km}$ diverge near the critical point with $V_{km}\sim(\lambda_{+}-\lambda)^{-1}(\lambda_{-}-\lambda)^{-1}$, where  $\lambda_{\rm -}(\kappa)=\sqrt{\tilde{\gamma}^{2}+(1-\tilde{\kappa})^{2}}/\sqrt{1-\tilde{\kappa}}$, see Appendix \ref{CMelements}. Moreover all covariant matrix elements are \emph{independent} on the parameters we wish to estimate which leads to significant simplification of the QFIM elements. Indeed, we have \cite{Nichols2018,Safranek2015}
\begin{equation}
\mathcal{F}_{km}=(\partial_{q_{k}}\bold{d}^{\rm T})\bold{V}^{-1}(\partial_{q_{m}}\bold{d}).
\end{equation}
Assuming that $\lambda_{+}(\kappa)<\lambda_{-}(\kappa)$ the ultimate uncertainty of the joint estimation becomes
\begin{eqnarray}
&&\delta F^{2}(\kappa)\ge \frac{\omega^{2}}{8}\{4(1+\tilde{\kappa})(\lambda_{+}^{2}(\kappa)-\lambda^{2})+\lambda^{4}+\lambda^{2}Q(\kappa)\},\notag\\
&&\delta \chi^{2}(\kappa)\ge \frac{\omega^{2}}{8F^{2}}\{4(1+\tilde{\kappa})(\lambda_{+}^{2}(\kappa)-\lambda^{2})+\lambda^{4}-\lambda^{2}Q(\kappa)\},\label{un_kappa}
\end{eqnarray}
where $Q(\kappa)=(\lambda^{2}-2-2\tilde{\kappa})\cos(2\chi)+2\tilde{\gamma}\sin(2\chi)$. As a comparison first I consider the case with $\kappa=0$ which corresponds to two uncoupled quantum probes. As can be expected in that case the additional factor of $2$ in the denominator compared to Eq. (\ref{uncertainty}) appears due to the additivity of the QFIM.

In Fig. \ref{fig4}(a) I show the ratio between the variances (\ref{un_kappa}) and those obtained for $\kappa=0$, namely $\delta F(\kappa)/\delta F$ and $\delta \chi(\kappa)/\delta \chi$. As can be seen the hopping $\kappa<0$ improves simultaneously force and phase sensitivities. In Fig. \ref{fig4}(b) I plot the uncertainty in the estimation of $F$ and $\chi$ where the phase is set $\chi=\chi_{\rm opt}$ with $\tan(2\chi_{\rm opt})=2\tilde{\gamma}(\lambda^{2}_{+}(\kappa)-2-2\tilde{\kappa})^{-1}$.
For such phase and coupling $\lambda$ close to the critical coupling $\lambda_{+}(\kappa)$ the off-diagonal elements of the QFIM vanishes, $\mathcal{F}_{F\chi}\approx0$. Approaching the critical point both uncertainties scales according to $\delta F(\kappa)\sim\frac{\omega}{\sqrt{2}}\sqrt{\lambda_{+}(\kappa)-\lambda}$ and $\delta\chi(\kappa)\sim\frac{\omega}{2F}\lambda^{2}_{+}(\kappa)$ or respectively $\delta F(\kappa)\sim\frac{\omega}{2}\lambda^{2}_{+}(\kappa)$ and $\delta \chi(\kappa)\sim\frac{\omega}{\sqrt{2}F}\sqrt{\lambda_{+}(\kappa)-\lambda}$ for phase $\chi=\chi_{\rm opt}+\pi/2$. We see that for given $\lambda$ and because $\lambda_{ +}(\kappa)<\lambda_{\rm c}$ one can achieve better sensitivity for $F$ and $\chi$ compared to the sensitivity which is achieved by using two uncoupled quantum probes with $\kappa=0$.

Further, one can evaluate the sum of the uncertainties of the dimensionless quadratures $q$ and $p$. I find
\begin{equation}
\delta q^{2}(\kappa)+\delta p^{2}(\kappa)\ge \frac{1}{4\nu}\{4(1+\tilde{\kappa})(\lambda^{2}_{+}(\kappa)-\lambda^{2})+\lambda^{4}\}.\label{sum}
\end{equation}
Since the displacement acts simultaneously on the two modes the SQL requires $\delta q^{2}(\kappa)+\delta p^{2}(\kappa)\ge \frac{1}{\nu}$. Close to the dissipative phase transition $\lambda\rightarrow \lambda_{ +}(\kappa)$ we have $\delta q^{2}(\kappa)+\delta p^{2}(\kappa)\sim \frac{\lambda_{\rm c}^{4}(\kappa)}{4\nu}$. Hence in order to overcome the SQL we require that $\lambda^{2}_{ +}(\kappa)<2$.  Moreover, the minimal value of right side of inequality (\ref{sum}) is $\tilde{\gamma}^{2}$ and thus as long as $\gamma<\omega$ the two parameters displacement estimation can operate beyond the SQL.

\section{Conclusion}\label{C}

In summary, I have discussed quantum sensor based on dissipative phase transition for the estimation of two displacement parameters. Our quantum probe consists of lattice system of two-level atoms and bosonic modes which interact via dipolar coupling. The interplay between the dissipation of bosonic excitations and the driven dynamics leads to a non-equilibrium steady state which exhibits non-analytical behaviour at the critical coupling. I have examined the sensitivity of the two displacement parameters and show that thanks of the driven-dissipative dynamics one can achieve enhancement of the parameters estimation compared to the non-driven case. I have shown that close to the dissipative phase transition one can achieve significant improvement of the sensitivity of one of the parameters namely magnitude or the phase of the displacement. Moreover, I have shown that the total uncertainty of the two parameters displacement estimation can overcome the SQL.

%%%%%%%%%%%%%%%%%%%%%%%
\section*{Acknowledgments}

PAI acknowledges support by the ERyQSenS, Bulgarian Science Fund Grant No. DO02/3.
%%%%%%%%%%%%%%%%%%%%%%%%%%%%%%%%%
%%%%%%%%%%%%%%%%%%%%%%%%%%%%%%%%%
\appendix
%%%%%%%%%%%%%%%%%%%%%%%%%%%%%%%%%
%%%%%%%%%%%%%%%%%%%%%%%%%%%%%%%%%
\section{Derivation of the SLD operators}\label{SLD_App}
We begin considering the SLD operators
\begin{equation}
\hat{\mathcal{L}}_{a}=2\sum_{m,n=0}^{\infty}\frac{\langle\psi_{m}|\partial_{a}\hat{\rho}_{q}|\psi_{n}\rangle}{p_{m}+p_{n}}|\psi_{m}\rangle\langle\psi_{n}|,
\end{equation}
where $a=F,\chi$ the parameters which we wish to estimate. The non-equilibrium steady-state has a Gaussian form and can be written as $\hat{\rho}_{q}=\sum_{n}p_{n}|\psi_{n}\rangle\langle\psi_{n}|$ with eigenstates $|\psi_{n}\rangle=\hat{R}(\delta)\hat{D}(\alpha)\hat{S}(\xi)|n\rangle$ and eigenvalues $p_{n}=N_{\rm th}^{n}/(1+N_{\rm th})^{n+1}$ where $N_{\rm th}$ stands the average number of thermal excitations. We have
\begin{equation}
N_{\rm th}=\frac{1}{2}\sqrt{\frac{4(\lambda^{2}_{\rm c}-\lambda^{2})+\lambda^{4}}{4(\lambda^{2}_{\rm c}-\lambda^{2})}}-\frac{1}{2}.\label{TE}
\end{equation}
First consider $a=F$. Since the information of the force $F$ is encoded only in the displacement parameter $\alpha$ we obtain
\begin{eqnarray}
\hat{\mathcal{L}}_{F}&=&2(\partial_{F}\alpha)\beta(-r,\phi)\sum_{n=0}^{\infty}\sqrt{n+1}\frac{p_{n}-p_{n+1}}{p_{n}+p_{n+1}}|\psi_{n+1}\rangle\langle\psi_{n}|\notag\\
&&-2(\partial_{F}\alpha)\beta(-r,-\phi)\sum_{n=0}^{\infty}\sqrt{n}\frac{p_{n}-p_{n-1}}{p_{n}+p_{n-1}}|\psi_{n-1}\rangle\langle\psi_{n}|.
\end{eqnarray}
Using Eq. (\ref{TE}) we arrive to
\begin{equation}
\hat{\mathcal{L}}_{F}=\frac{2\partial_{F}\alpha}{1+2N_{\rm th}}\hat{R}\hat{D}\hat{S}\{\beta(r,\phi) \hat{a}^{\dag}+\beta(r,-\phi)\hat{a}\}\hat{S}^{\dag}\hat{D}^{\dag}\hat{R}^{\dag},
\end{equation}
where $\beta(r,\phi)=\cosh(r)+e^{2i\phi}\sinh(r)$.

Next we consider the SLD operator $\hat{\mathcal{L}}_{\chi}$. Now the information of the phase is encoded in $\delta$, $\alpha$ and $\phi$. Thus the derivatives of $\hat{\rho}_{q}$ contains three terms, namely
\begin{equation}
\hat{\mathcal{L}}_{\chi}=\hat{\Lambda}_{\delta}+\hat{\Lambda}_{\alpha}+\hat{\Lambda}_{\phi}.\label{L}
\end{equation}
Here
\begin{equation}
\hat{\Lambda}_{\delta}=2i\partial_{\chi}\delta\sum_{n,m=0}^{\infty}\frac{p_{n}-p_{m}}{p_{n}+p_{m}}\langle\psi_{m}|\hat{a}^{\dag}\hat{a}|\psi_{n}\rangle|\psi_{m}\rangle\langle\psi_{n}|.
\end{equation}
The expression can be written as
\begin{eqnarray}
\hat{\Lambda}_{\delta}&=&2i\frac{(\partial_{\chi}\delta)\alpha}{1+2N_{\rm th}}\hat{R}\hat{D}\hat{S}\{\beta(-r,\phi) \hat{a}^{\dag}-\beta(-r,-\phi)\hat{a}\}\hat{S}^{\dag}\hat{D}^{\dag}\hat{R}^{\dag}\notag\\
&&-i\sinh(2r)\frac{(\partial_{\chi}\delta)(1+2N_{\rm th})}{1+2N_{\rm th}+2N_{\rm th}^{2}}\hat{R}\hat{D}\hat{S}(e^{2i\phi}\hat{a}^{\dag2}\notag\\
&&-e^{-2i\phi}\hat{a}^{2})\hat{S}^{\dag}\hat{D}^{\dag}\hat{R}^{\dag}.\label{L1}
\end{eqnarray}

The second term in (\ref{L}) arrives due to the derivative of the displacement amplitude $\alpha$ with respect to $\chi$. We have
\begin{equation}
\hat{\Lambda}_{\alpha}=\frac{2\partial_{\chi}\alpha}{1+2N_{\rm th}}\hat{R}\hat{D}\hat{S}\{\beta(r,\phi) \hat{a}^{\dag}+\beta(r,-\phi)\hat{a}\}\hat{S}^{\dag}\hat{D}^{\dag}\hat{R}^{\dag}.\label{L2}
\end{equation}
Finally, we can evaluate the last term in (\ref{L}) by using the expression
\begin{equation}
\partial_{\chi}e^{-i \hat{A}}=-i\int_{0}^{1}ds e^{-i \hat{A}}e^{is \hat{A}}(\partial_{\chi}\hat{A})e^{-is \hat{A}},
\end{equation}
where $\hat{A}$ is hermitian operator. Then we obtain
\begin{equation}
\hat{\Lambda}_{\phi}=-i\sinh(2r)\frac{(\partial_{\chi}\phi)(1+2N_{\rm th})}{1+2N_{\rm th}+2N_{\rm th}^{2}}\hat{R}\hat{D}\hat{S}(e^{2i\phi}\hat{a}^{\dag2}
-{\rm h.c.})\hat{S}^{\dag}\hat{D}^{\dag}\hat{R}^{\dag}.\label{L3}
\end{equation}
Combining Eqs. (\ref{L1}), (\ref{L2}), and (\ref{L3}) in Eq. (\ref{L}) and using that $\partial_{\chi}(\delta+\phi)=0$ we find
\begin{equation}
\hat{\mathcal{L}}_{\chi}=\frac{2}{1+2N_{\rm th}}\hat{R}\hat{D}\hat{S}(\upsilon \hat{a}^{\dag}+\upsilon^{*}\hat{a})\hat{S}^{\dag}\hat{D}^{\dag}\hat{R}^{\dag},
\end{equation}
with $\upsilon=(\partial_{\chi}\alpha)\beta(r,\phi)+i\alpha(\partial_{\chi}\delta)\beta(-r,\phi)$.

Next we evaluate the commutator between the two SLD operators. We have
\begin{equation}
[\hat{\mathcal{L}}_{F},\hat{\mathcal{L}}_{\chi}]=\frac{4\partial_{F}\alpha}{(1+2N_{\rm th})^{2}}\{\beta(r,-\phi)\upsilon-\beta(r,\phi)\upsilon^{*}\}\hat{\bold{1}}.
\end{equation}
Hence the commutator coincide with the average one. Further we can simplify to
\begin{equation}
[\hat{\mathcal{L}}_{F},\hat{\mathcal{L}}_{\chi}]=8i\frac{\alpha(\partial_{F}\alpha)(\partial_{\chi}\delta)}{(1+2N)^{2}}\hat{\bold{1}}.
\end{equation}
Finally, we obtain
\begin{equation}
[\hat{\mathcal{L}}_{F},\hat{\mathcal{L}}_{\chi}]=\frac{F}{\omega^{2}}\frac{8i}{4(\lambda^{2}_{\rm c}-\lambda^{2})+\lambda^{4}}\hat{\bold{1}},
\end{equation}
which emphasize that the commutator is independent on the phase.
\section{Derivation of the QFIM elements}\label{SLD_Operators}

Having in hand the SLD operators one can evaluate the QFIM elements using
\begin{equation}
\mathcal{F}_{km}=\frac{1}{2}{\rm Tr}(\hat{\rho}_{q}\{\hat{\mathcal{L}}_{q_{k}},\hat{\mathcal{L}}_{q_{m}}\}).
\end{equation}
We find that the diagonal elements becomes
\begin{equation}
\mathcal{F}_{F}=\frac{4(\partial_{F}\alpha)^{2}}{1+2N_{\rm th}}|\beta(r,\phi)|^{2},\quad \mathcal{F}_{\chi}=\frac{4}{1+2N_{\rm th}}|\upsilon|^{2},
\end{equation}
which can be rewritten as
\begin{eqnarray}
&&\mathcal{F}_{F}=\frac{4(\lambda^{2}_{\rm c}-\lambda^{2})+\lambda^{4}-\lambda^{2}Q}{\omega^{2}(\lambda^{2}_{\rm c}-\lambda^{2})(4(\lambda^{2}_{\rm c}-\lambda^{2})+\lambda^{4})},\notag\\
&&\mathcal{F}_{\chi}=\tilde{F}^{2}\frac{4(\lambda^{2}_{\rm c}-\lambda^{2})+\lambda^{4}+\lambda^{2}Q}{(\lambda^{2}_{\rm c}-\lambda^{2})(4(\lambda^{2}_{\rm c}-\lambda^{2})+\lambda^{4})},
\end{eqnarray}
with $Q=\{(\lambda^{2}-2)\cos(2\chi)+2\tilde{\gamma}\sin(2\chi)\}$. Both elements diverges approaching the critical coupling $\lambda_{\rm c}$. The latter implies that for a single parameter estimation where only one of the parameters is estimated the respective sensitivity is enchanted closed to the dissipative phase transition.

The off-diagonal elements of the QFIM describe the correlation between the two parameters. We find
\begin{equation}
\mathcal{F}_{F\chi}=\frac{2\partial_{F}\alpha}{1+2N_{\rm th}}\{\beta(r,-\phi)\upsilon+\beta(r,\phi)\upsilon^{*}\},
\end{equation}
which can be rewritten as
\begin{equation}
\mathcal{F}_{F\chi}=\frac{F\lambda^{2}}{\omega^{2}}\{(\lambda^{2}-2)\sin(2\chi)-2\tilde{\gamma}\cos(2\chi)\}.
\end{equation}
The correlation vanishes for the optimal phase given by $\tan(2\chi_{\rm opt})=2\tilde{\gamma}/(\lambda^{2}-2)$.

\section{Covariant Matrix for two mode Gaussian state}\label{CMelements}
Here we provide information for the symmetric covariant matrix elements for the two mode steady state Gaussian state. We find that the diagonal elements are
\begin{eqnarray}
&&V_{11}=A^{-1}\{2\tilde{\gamma}^{4}+\tilde{\gamma}^{2}(4\tilde{\kappa}^{2}+(4-3\lambda^{2}))\notag\\
&&\quad\quad+(\tilde{\kappa}^{2}-1)(2\tilde{\kappa}^{2}-(2-3\lambda^{2}+\lambda^{4}))\},\notag\\
&&V_{22}=A^{-1}\{2\tilde{\gamma}^{4}+(\tilde{\kappa}+1-\lambda^{2})(\tilde{\kappa}-1+\lambda^{2})(2\tilde{\kappa}^{2}+\lambda^{2}-2)\notag\\
&&\quad\quad+\tilde{\gamma}^{2}(4\tilde{\kappa}^{2}+(4-5\lambda^{2}+\lambda^{4}))\},
\end{eqnarray}
and $V_{33}=V_{11}$, $V_{44}=V_{22}$. Here we have defined $A=2(\tilde{\kappa}^{2}-\kappa^{2}_{-})(\tilde{\kappa}^{2}-\kappa^{2}_{+})$. The off-diagonal elements are given by
\begin{eqnarray}
&&V_{12}=A^{-1}\tilde{\gamma}\lambda^{2}(\tilde{\gamma}^{2}+\tilde{\kappa}^{2}-\lambda^{2}+1),\notag\\
&&V_{13}=A^{-1}\tilde{\kappa}\lambda^{2}(\tilde{\gamma}^{2}+\tilde{\kappa}^{2}-1),\notag\\
&&V_{24}=A^{-2}\tilde{\kappa}\lambda^{2}\{(1-\lambda^{2})^{2}-(\tilde{\gamma}^{2}+\tilde{\kappa}^{2})\},\notag\\
&&V_{14}=A^{-2}\tilde{\gamma}\tilde{\kappa}\lambda^{2}(\lambda^{2}-2),
\end{eqnarray}
with $V_{34}=V_{12}$ and $V_{23}=V_{14}$. Note that the factor $A$ can be rewritten as follows
\begin{equation}
A=(1-\tilde{\kappa}^{2})(\lambda^{2}_{+}(\kappa)-\lambda^{2})(\lambda^{2}_{-}(\kappa)-\lambda^{2})
\end{equation}
and thus all elements diverge near the critical point with $V_{km}\sim(\lambda_{+}-\lambda)^{-1}(\lambda_{+}-\lambda)^{-1}$.

Since the covariant matrix elements are independent on the parameters we wish to estimate such that the expression for the QFIM elements is given by $\mathcal{F}_{km}=(\partial_{q_{k}}\bold{d}^{\rm T})\bold{V}^{-1}(\partial_{q_{m}}\bold{d})$. We find that diagonal elements are
\begin{eqnarray}
&&\mathcal{F}_{F}=\frac{2\{4\tilde{\gamma}^{2}+((\lambda^{2}-2)-2\tilde{\kappa})-
\lambda^{2}Q(\kappa)\}}{\omega^{2}(\tilde{\kappa}-\kappa_{+})(\tilde{\kappa}-\kappa_{-})(4\tilde{\gamma}^{2}+((\lambda^{2}-2)-2\tilde{\kappa})^{2})},
\notag\\
&&\mathcal{F}_{\chi}=\frac{2\tilde{F}^{2}\{4\tilde{\gamma}^{2}+((\lambda^{2}-2)+2\tilde{\kappa})-
\lambda^{2}Q(\kappa)\}}{(\tilde{\kappa}-\kappa_{+})(\tilde{\kappa}-\kappa_{-})(4\tilde{\gamma}^{2}+((\lambda^{2}-2)-2\tilde{\kappa})^{2})},
\end{eqnarray}
where $Q(\kappa)=\{((\lambda^{2}-2)-2\tilde{\kappa})\cos(2\chi)+2\tilde{\gamma}\sin(2\chi)\}$. The off-diagonal element are
\begin{equation}
\mathcal{F}_{F\chi}=\frac{2F\lambda^{2}\{((\lambda^{2}-2)-2\tilde{\kappa})\sin(2\chi)-2\tilde{\gamma}\cos(2\chi)\}}
{\omega^{2}(\tilde{\kappa}-\kappa_{+})(\tilde{\kappa}-\kappa_{-})(4\tilde{\gamma}^{2}+((\lambda^{2}-2)-2\tilde{\kappa})^{2})}.
\end{equation}
%%%%%%%%%%%%%%%%%%%%%%%%%%%%%%%%%%%%%%%%%%%%%%%%%%%%%%%%%%%%%%%%%%%%%%%%%%%%%%%%%%%%%%%%%%%%%%%%%%%%%%
%%%%%%%%%%%%%%%%%%%%%%%%%%%%%%%%%%%%%%%%%%%%%%%%%%%%%%%%%%%%%%%%%%%%%%%%%%%%%%%%%%%%%%%%%%%%%%%%%%%%%%

\end{document}